# Information and communication technology initiatives for knowledge sharing in agriculture

SIDDHARTHA PAUL TIWARI

*Google Inc*

## ABSTRACT

A survey on status and trends of information and communication technologies (ICT) use for knowledge sharing in agriculture was attempted. Among asian countries, India comes under the second next category after the advanced user category comprising Japan, South Korea and Taiwan. Both profit-motive and business auginentation on one hand and community services and rural welfare on the other have been the objectives of ICT-based models in agriculture in India. Major ICTs endeavours for agriculture and rural concerns include multifaceted projects and services of National Informatics Centre, *"Bhoomi"* of Karnataka, Village Information Centres of Government of India, ITC's e-Choupal, TARAhaat, iKisan.com, i-Villages of Puducherry (formerly Pondicherry) of Swaminathan Research Foundation, i-Community of Hewlett-Packard, Warana Wired Village Project, n-Logue, 'Gyandoot of Government of Madhya Pradesh, Knowledge Network for Grass Root Innovations of the Society for Research and Initiatives for Sustainable Technologies and Institution, Traditional Knowledge Digital Library and several others. The ICT endeavours for agriculture belong to a wide array of agencies, viz private sector, public sector, self-help groups and NGOs, and also include combined endeavours, e-Learning is being increasingly resorted to both in (i) in campus or 'presence mode, and (ii) distance' mode. Its use is gradually easing-out the stakeholders from the stranglehold of the inter-deterrence of the 3 arms of the 'Iron Triangle', viz (i) quality, (ii) access, and (iii) cost. The social groups having less mobility are poised to benefit more from this mode of education. This could also be one of the potent tools to bring about gender mainstreaming. e-Learning is being integrated into the existing organizational and educational structure as a hybrid system that can be called 'ICT-supported learning'. Connectivity, content development, infrastructure development, faculty development, need assessment on a continuum, linking the nodes and formation of consortia etc. are the areas identified that need to be supported and developed.

Knowledge is an increasingly significant factor of system productivity and value chains in agriculture. The report of the Task Force on India as Knowledge Superpower (GOI 2001) has also emphasized the necessity of developing the capacity to generate, absorb, disseminate and protect knowledge and exploit it as a powerful tool to derive societal transformation. The contribution of Information and Communication Technology (ICT) to the national development including agricultural development has been steadily increasing since early 1990s'. In ICT-enabled network, everything is added up. When the information is networked, the power and utility of the information grows tremendously. Metcalfe's law (Gilder 1993, Metcalfe 1995) states that the value of a telecommunication network is proportional to the square of the number of users of the system. There is annihilation of distance and near instantaneous observation of events wherein real time response can be ensured and owing to this feature the recording and replaying cannot substitute a telecommunication network. The costs incurred are low and benefits are phenomenal. Further, the computational resources can be accessible and connected countrywide so that machines can talk to each other.

India presently has networks of over 700 000 route km of optical fibre cables of various sizes spread almost all over the country (Seth 2006). The major telecom service providers offerings data network/bandwidth are Bharat Sanchar Nigam Ltd, Videsh Sanchar Nigam Ltd, Reliance, Airtel, Railtel and Power Grid along with Government networks, viz NICNET of National Informatics Centre of the Department of Information Technology and State Wide Area Network by almost all the states and UTs. Further, there are dedicated

Product Specialist, Google India Pvt. Ltd and PhD Scholar (E-mail: siddhartha@google.com), Department of Management Studies, Indian Institute of Technology, Delhi network for research and education community, such as Education and Research Network (ERNET) and UGC INFONET; the latter has been overlaid on ERNET infrastructure to provide assured quality of service and optimal utilization of bandwidth resources. It has a scalable architecture to grow from universities to affiliated colleges. Several of these national networks are connected to global networks. The major global networks are Internet 2 (a US networking consortium), GÉANT2 (Pan European Network), US National Lambda Rail, and ESnet (US DoE's Energy Services net) among several others. These networks have acted as a meeting place for educational and developmental institutions alike (Seth 2006).

Of late, significant endeavours have been witnessed in using and promoting knowledge sharing in agriculture. The use of ICT has gone beyond simply establishing information flow channels and has started integrating itself with meeting the various livelihood needs (natural, social, human, physical and financial) of the rural community. The present study attempts to analyze some basic elements of the phenomenon and takes stock of some of the model sample cases in Indian agriculture in this regard.

The methodology of the study mainly comprised survey of literature, obtaining responses on a structured questionnaire and personal/telephonic interviews including telephonic follow-up towards response on structured questionnaire. The paper is organized into sections, viz the terms explained, e-learning, m-learning, status of ICT use in agricultural R&D, major ICT endeavours for agriculture in India, intellectual property rights issues, and suggestions for future strategy.

## INFORMATION TECHNOLOGY, INFORMATION COMMUNICATION TECHNOLOGY AND

E-LEARNING - THE TERMS Information technology (IT), as defined by the Information Technology Association of America (ITAA), is the study, design, development, implementation, support or management of computer-based information systems, particularly software applications and computer hardware. In short, IT deals with the use of electronic devices, such as computers (including mobile ones and those in cellphones) and computer software to convert, store, protect, process, transmit and retrievory information, securely. Information science encompasses several areas of basic and applied research. Mathematical logic, theory of computation, quantum computing, data encryption, parellel alogrithms, random alogrithms etc. are some areas of theoretical nature on one hand and the computer network, digital electronics, simulation, graphics, virtual reality, artificial intelligence, biocomputing etc, are areas of applied nature on the other hand.

Recently it has become popular to broaden the term IT to ICT (Information and Communication Technology) to include the field of electronic communication. Hence the ICT is, therefore, an umbrella term that includes any communication device or application, encompassing electronic communication. It includes use of radio, television, cellular phones, computer and network hardware and software, satellite systems and so on, as well as the various services and applications associated with them, such as videoconferencing and distance learning. ICTs are often spoken of in a particular context, such as ICTs in education, health care or libraries.

Terms, viz 'distributed education' and 'e-learning' also need to be explicitly understood. Distributed education is independent of fixed time and place, and delivers course content online to distant, commuting and residential students alike. Distance learning is a subset of distributed or continuous education. Distance education is planned learning that normally occurs in a different place from teaching and as a result requires special techniques of the course design, special instructional techniques, special methods of communication by electronic and the other technology, as well as special organizational and administrative arrangements. It resorts to e-Learning as an approach and ICTs as tools of content delivery.

The term 'e-learning is defined by the Commission of the European Communities as the use of new multimedia technologies and the internet to improve the quality of learning by facilitating access to resources and services as well as remote exchanges and collaboration (CEC 2001). Online learning refers more specifically to the context of using the Internet and associated web-based applications as the delivery medium for the learning experience. e-Learning is flexible learning involving the use of computer, internet/world wide web, CD-ROM, software, media and the other ICT resources. e-Learning comes in many variations and often a combination of these variations, such as (i) purely online--no face-to-face meetings, (ii) blended learning-- combination of online and face-to-face, (iii) synchronous or asynchronous, (iv) instructor-led group, (v) self-study with or without subject matter expert, (vi) web-based, (vii) computer-based (CD-ROM) and (viii) video/audio tape etc.

e-Agriculture goes beyond technology to promote the integration of technology with multimedia, knowledge and culture, with the aim of improving communication and learning processes between various sectors in agriculture locally, regionally and worldwide. Facilitation, support of standards and norms, technical assistance, capacity building, education, and extension are all the key components to e Agriculture. Furthermore, e-Agriculture objectives include improving the effectiveness of traditional communication channels and already existing communication practices.

STATUS OF ICT USE IN AGRICULTURE WITH PARTICULAR REFERENCE TO R&D IN ASIA ICTs have encouraged a teaching and learning milieu that recognizes that people both across and within the countries operate differently, have different learning styles and have culturally diverse perspectives. In regard to the use of ICT in agricultural research and development, Maru (2003) has categorized the Asia-Pacific countries into 4 groups as given below,

Maru (2003) has ascribed uneven or heterogenic use of ICT in agricultural R&D in Asia to (i) lack of capacity, (ii) NARS leadership, (iii) funding of ICT in NARS, (iv) infrastructure, (v) skills to use and manage ICT and Information, (vi) lack of appropriate technologies and models to use ICT for ARD, and (vii) lack of clear policies. Use of ICT in agricultural R&D is dependant not only on agricultural research policies but a wide envelope of policies related to (a) rural telecommunications, (b) rural development, (c) Agricultural development, (c) infrastructure, especially electricity, (d) education and (e) governance. Common issues in ICT use in agricultural R&D, as identified by Maru (2003) are prioritization of areas and focus on activities to be taken up. In view of financial and skill constraints of the developing countries, the focus of ICT use could be on (i) scientific and technical information, (ii) research management and research data management, (iii) extension and outreach, (iv) agricultural education, (vi) enabling communication and messaging between institutions and/or researchers (Maru 2003).

(i) quality, (ii) access and (iii) cost. The 3 facets constitute the 'Iron Triangle'. Positive change in one facet of the iron triangle may affect the other(s) adversely. Stranglehold of the iron triangle can be broken by use of E-learning. Besides, E-learning also facilitates distributed education and continuous education or life-long learning. E-learning could be used both in (i) in campus or "presence' mode, and (ii) 'distance' or distributed education mode. A survey of literature and some interviews conducted by the author have shown that there is no general or exclusive move from campus learning to e-Learning both in "presence and 'distance modes, but rather a more selective behaviour is observed that can be labelled as integrated or hybrid system. This does not mean that e-Learning is not spreading, e-Learning is undoubtedly spreading in both 'presence and 'distance mode but is not being solely resorted to as an exclusive approach rather it is being integrated into the existing organizational and educational structure as a hybrid system that can be called 'ICT-supported learning'. The pace of integration in favour of e-learning is poised to become increasingly rapid. Distance and Open Education, in particular, could be used to reach the unreached; almost anytime anywhere in less time and less cost. The social groups having less mobility are poised to benefit more from this mode of education. In India and other countries of the Indian sub-continent, this could be one of the potent tools to bring

E-LEARNING IN

AGRICULTURAL EDUCATION

about gender mainstreaming as well. Providing education to Given India's strengths in the
ICT and software sectors, all the needy and closing the digital divide will depend not
e-learning is bound to increase in scope and utility. An array only on technology but also
on providing the skills and of innovations and experiments are underway to overcome
content that is most beneficial. the educational challenges. India has launched its own
Determining and enforcing quality standards is as educational satellite called 'Edusat for
linking schools across controversial and elusive for distributed education as it is the
country to mitigate the issue of absence of qualified for higher education in general. India
already has Distance teachers, as well as the lack of access to schools. Several Education
Council (DEC) functioning under the aegis of the key initiatives, such as the ICT@ schools
programme and Indira Gandhi National Open University. There is, however, the Vidya
Vahini Project have also been launched to train a need is felt to make DEC an independent
entity. A National children in frontline technologies, as well as to improve the Mission for
Education through ICT has been proposed by efficacy and spreading of teaching. Various
State the Human Resource Ministry under which all institutions Governments have also
launched path-breaking programmes of higher learning would be networked through
broadband for enabling remote access, improving pedagogic content, connectivity,
e-Content would be developed and made and facilitating community involvement in
teaching. available through the medium of Edusat, Internet and cable EDUSAT
educational programme launched in Haryana, TV networks. The National Programme on
Technology ISRO's initiative to link Delhi's engineering colleges and Enhanced Learning
(NPTEL) is also promoting the Virtual Schools and Learning Home project in Maharashtra
development of e-learning resources. are examples of this type.

    If the trend in West in general, and Europe in particular, India has also been partnering
with countries in Africa in is to be taken as an indicator, then the learning is moving the
area of ICT-enabled learning, and there is considerable from being "teacher-centric' to
'learner-centric' or 'student potential for such international collaborative ventures
centric' (Succi and Cantoni 2005). To attract students and to (Nambiar 2005).
be in a utilitarian mode, the use of computers and ICTs has E-Learning is being increasingly
resorted to in higher to be necessarily resorted to. A survey carried out by an e education.
Although the main reason for this appears to be learning solutions provider, WebCT,
highlighted that an objective of providing wider access, this needs some universities now
see e-Learning as 'mission critical and analysis. There are 3 main facets in imparting
education, viz offering greater access to better quality education. Three

# MAJOR ICTS ENDEAVOURS FOR AGRICULTURE AND RURAL CONCERNS IN INDIA

quarters of the 150 institutions surveyed stated that computer based learning had a major role in most of their courses or they would do so within 3 years.

The case of UK's 'e-university', 'UKeU', is an interesting one. The university has failed. Its failure created a furore. The surveys, however, indicate a growing demand for online courses in spite of the failure of the UKeU. Part of UKeU's problem was that students preferred to work through existing universities of established reputations, which have been developing their own e-Learning materials. The choice was probably between the reputed and a new university rather than between traditional and e-Learning approaches. If reputed universities use e-Learning, it will have very high acceptance.

Use of ICT has been made by several agencies in India, both public and private, in agriculture for promoting the sale of products of a company, transfer of technology etc. (Swaminathan 1993, Meera 2002, Meera *et al.* 2004, Rao 2007). Several of them are participatory in nature as well. Many initiatives have been undertaken in India using ICT for agricultural development. Some major endeavours of knowledge sharing in agriculture, where ICTs have been used successfully are categorized based on their objectives and use, and are analyzed as given below.

*m-Learning*

By mixing e-Learning and mobile computing, a new form of education may be created. It is called mobile learning or "m-learning'. Although related to e-Learning and distance education, it is yet distinct in its focus on learning across contexts and learning with mobile devices. One definition of mobile learning is: 'Learning that happens across locations, or that takes advantage of learning opportunities offered by portable technologies.' The term covers (1) learning with portable technologies, where the focus is on the technology (which could be in a fixed location, such as a classroom); (ii) learning across contexts, where the focus is on the mobility of the learner, interacting with portable or fixed technology; and (iii) learning in a mobile society, with a focus on how society and its institutions can accommodate and support the learning of an increasingly mobile population.

With General Packet Radio Service (GPRS) connectivity improving and charges coming down, m-Learning is bound to increase. All one need to do is download the software from an appropriate site. For example, sites, such as Wizdom.in and likes of them offer a Graduate Record Examination (GRE) training programme that can be downloaded to a GPRS mobile phone. All this is for a price far less than what most coaching institutes charge. Students in India, for example those in Chennai, have already started with m

Learning. Mobile learning is the next frontier as one segment of the market is ready to move beyond voice and entertainment to explore the other dimensions of this handheld computer.

Most personal technologies can support mobile learning, including (i) Personal Digital Assistant, in the classroom and outdoors, (ii) Tablet PC, Ultra-Mobile PC (UMPC), mobile phone, camera phone and smartphone, (iii) personal audio player, e g for listening to audio recordings of lectures, and (iv) handheld audio and multimedia guides in museums and galleries.

Several Krishi Vigyan Kendras (KVKs) of the ICAR have already started "Kisan Mobile Sandesh' service which provides needed knowledge to farmers.

*Multifaceted projects and services of National Informatics Centre*

Using ICTs in agriculture and allied services, National Informatics Centre (NIC) has undertaken many projects mainly for the central Department of Agriculture and Cooperation, GoI, to provide relevant agricultural information in rural areas, helping farmers to improve their labour productivity, increase their yields, and realize a better price for their produce (NIC 2005). These include (i) AGRISNET-an infrastructure network up to block level agricultural offices facilitating agricultural extension services and agribusiness activities to usher in rural prosperity, (ii) AGMARKNET--With a road map to network 7 000 Agricultural produce wholesale markets and 32 000 rural markets, (iii) ARISNET--Agricultural Research Information System Network, (iv) SeedNET-Seed Informatics Network, (v) CoopNet-to network 93 000 Agricultural Primary Credit Societies (PACS) and Agricultural Cooperative Marketing Societies to usher in ICT enabled services and rural transformation, (vi) HORTNET Horticultural Informatics Network, (vii) FERTNET Fertilisers (chemical, bio and organic manure) Informatics Network facilitating Integrating Nutrient Management' at farm level, (viii) VISTARNET ---Agricultural Extension Information System Network, (ix) PPIN-Plant Protection Informatics Network, (x) APHNET-Animal Production and Health Informatics Network networking about 42 000 animal primary health centres, (xi) FISHNET._Fisheries Informatics Network, (xii) LISNET-Land Information System Network linking all institutions involved in land and water management for agricultural productivity and production systems, which has now evolved as 'Agricultural Resources Information System' project being implemented through NIC, (xiii) AFPINET-Agricultural and Food Processing Industries Informatics Network, (xiv) ARINET Agricultural and Rural Industries Information System Network to strengthen Small and Micro Enterprises (SMEs), (XV) NDMNET--Natural Disaster Management Knowledge Network, and (xvi) Weather NET_Weather Resource System of India. 6)

*Providing authenticated land records and information*
*Bhoomi, Karnataka:* Government of Karnataka initiated *Bhoomi* (meaning "land') in 2001 to document rather computerize the land records of farmers. Around 20 million land records of 6.7 million landowners in 176 talukas (27 000 villages) of Karnataka have been computerized and RTC kiosks have been opened in those talukas' using the 'Bhoomi'. Farmers have albeit to pay for it but nominally. They visit the kiosks and obtain their updated land records for a user charge of Rs 15 by spending only 5–30 min. of their time. A World Bank study has assessed that farmers of Karnataka now save about 1 billion rupees (Orbicom 2004). About 66% of farmers used kiosks with no help and most of them (78%) found the system to be very simple (Lobo and Balakrishnan 2002).

*Village Information Centres of GoI*
The National Institute of Agricultural Extension Management (MANAGE) of the Ministry of Agriculture, Government of India started a 'cyber Extension Programme' in October 2000 to facilitate farmers' access to agricultural information. The programme provides internet connectivity at district level (24 districts) in 7 states, viz Andhra Pradesh, Bihar, Himachal Pradesh, Jharkhand, Maharashtra, Orissa and Punjab. The model has been extended to village level in Andhra Pradesh. Village Information centres operate in 11 villages in Rangareddy district of Andhra Pradesh each centre extending facilities to surrounding 25-30 villages and about 20 000 to 30 000 farmers. The village information centres are located at Mutually Aided Cooperative Thrift and Credit Society (MACTIC). In Andhra Pradesh, 45 000 people have already benefited from this initiative in 300 villages of 10 blocks.

> the sale of farm inputs and purchase farm produce from the farmers' doorsteps. Real-time information and customized knowledge provided by 'e-Choupal' enhance the ability of farmers to take decisions and align their farm output with market demand and secure quality and productivity.

e-Choupal/-soya-Choupal' serves primarily as a direct marketing channel, virtually linked to the 'M*andi s*ystem'. It eliminates many intermediaries (Adhiguru and Mruthyunjaya 2004) and, thus, improves sale value realized by the farmers for the produce (about 2%). ITC gains through a better control over quality of produce and reduction in procurement cost by 2.5% (Rs 250/tonne at current prices). It is estimated that there was mark-up of 7-8% on the price of soybean from the farm gate to factory gate. Of this about 2.5% was borne by the farmer and 5% was bome by ITC. Farmers selling directly to ITC through 'e-Choupals' saved about Rs 120/tonne on transaction costs and gained access to good quality inputs and knowledge, while the Company also saved about Rs 215/tonne on its transaction costs and had access to good quality produce for its trade and processing industry (Rao 2007).

The problems encountered while setting up and managing these "e-Choupals' are primarily of infrastructural inadequacies, including power supply, telecom connectivity and bandwidth, apart from the challenge of imparting skills to the first time internet users in remote and inaccessible areas of rural India. The ITC has provided several alternative and innovative solutions to overcome these challenges such as power back-up through batteries charged by solar panels, upgrading BSNL exchanges with RNS (Radio Network Subsystem) kits, installation of VSAT equipment, Mobile Choupals, local caching of static content on website to stream in the dynamic content more efficiently, 24 x 7 helpdesk etc.

The initiative has got extended to several crops now. 'e Choupal' is claimed to have already become the largest initiative among all Internet-based interventions in rural India. 'e-Choupal' services today reach out to more than 4 million farmers growing a range of crops - soybeans, coffee, wheat, rice, pulses, shrimp-in some 40 000 villages through almost 6 500 kiosks across 8 states, viz Madhya Pradesh, Maharashtra, Rajasthan, Haryana, Karnataka, Andhra Pradesh, Uttar Pradesh and Uttaranchal.

*TARAhaat:* TARAhaat.com was started in 1999 by an NGO. It is a business enterprise of Development Alternatives (DA) that focuses on sustainable rural development in India (mainly Uttar Pradesh and Punjab). The DA has 'Technology and Action for Rural Advancement (TARA)' as its marketing arm. Its objective is to improve information flows, computer literacy, crop and market information. The business model combines a mother portal, TARAhaat.com, with a network of 18 franchised village Internet centres, or TARAkendras in Uttar Pradesh and Punjab. These kendras deliver computer based education, information, services, and online market opportunities. Telephone connectivity and VSAT links are

*Facilitating marketing and input availability*

*ITC's e-Choupal:* The idea of establishing the web-based initiative, "e-Choupal', was mooted and implemented by the international business division of Indian Tobacco Company (ITC) in 2000. "*Choupal* is a familiar Hindi word, particularly in the rural masses of north India and denotes a community place where villagers sit and talk leisurely after the end of day's work. ITC implemented 'e-Choupal (also called "soy-Chaupal') first for soybean crop in Madhya Pradesh in India. The prime objective was to enhance efficiency in ITC's procurement of soybean. The network covers about 1 200 *Choupal*s at the village level in Madhya Pradesh. The *Choupals* are provided with internet connectivity with solar panel battery back up and VSAT (Very Small Aperture Terminal, a satellite communications system) equipment. Village internet kiosks managed by farmers, called '*sanchalaks*", themselves, enable the agricultural community's access to information in their local language on the weather and market prices, disseminate knowledge on scientific farm practices and risk management, facilitate being used for networking internet centers. Content is mainly in English and Hindi and likely to extended in other regional languages.

*iKisan.com:* It is an initiative of the Nagarjuna Group of private companies. It is a platform for providing/sale of its own and their party products (farm inputs), services (knowledge-based crop management) and information (weather, markets) to meet farmers' agricultural needs. It maintains a web portal in local language at the company site and an information kiosk staffed by its trained representatives in villages. The main features of the *iKisan* business model are aggregation of demand to enable supply of good quality inputs at relatively low prices and provision of technical advice to farmers through the portal, the kiosk and its representatives. It operates mainly in Andhra Pradesh and Tamil Nadu. It really proved popular in Andhra Pradesh where 9 technical centres (kiosks) were established in different districts (Meera *et al.* 2004). Besides *iKisan*.com website, there are technical centres that are operated mainly by agricultural graduates, linking the website and the farmers for facilitated access. Project services are available only to member farmers. Farmers can become members by payment of a nominal fee. The aim is to enhance agricultural productivity, mainly of crops. Information on input availability, market information, disease and insect-pests management etc. is provided. *Ikisan* also provides interaction between farmers and experts to solve specific problems and answer the queries through a service component called 'Let US Talk' that comprises Chat, Expert Chat, Bulletin and Ask Us.

Information Centre (CIC) set up in the village that provides the interface for assembling an ecosystem of public-private partnerships to accelerate development while opening new markets and developing new products and services. HP provides for reliable ICT Infrastructure for broadband access at the CIC through VSAT to resolve last mile connectivity issues. It also invests in capacity building of operators and networking with a range of public and private organizations through the CIC. HP is also creating intangible business assets (eg., new networks and increased familiarity with new markets). The experiment has already created business value to HP. Many of the online products developed at Kupam have potential for application in other areas in Asia and Africa where HP has set up similar CICs.

*Warana Wired Village Project:* The project demonstrates the use of ICT towards accelerated socio-economic development of villages around Warana Nagar in the

Kolhapur and Sangli districts of Maharashtra. The project area has a cluster of 70 villages, consisting of 46 villages from Kolhapur and 24 villages from Sangli district. It has been jointly implemented by National Informatics Centre (NIC) of GoI, Directorate of Information Technology, Government of Maharashtra and the Warana Sahakari Dudh Utpadan Prakriya Ltd, Warana Nagar, Maharashtra (WSDUPL). The project utilizes IT to increase the efficiency and productivity of the existing co-operative enterprise by setting up a state-of-art computer communication network. The project is led by the Warana co-operative sugar factory, The Warana co-operative complex in Maharashtra has become famous as a fore-runner of successful integrated rural development emerging from the co-operative movement. The project was initiated with 6 business centres, 6 IT centres and 70 village booths (kiosks). The kiosk managers are employees of the Warana Nagar Co-operative Society. The project provides agricultural, medical, and educational information to villager at facilitation booths in their villages. It provides communication facilities at the booths to link villages to the Warana co-operative complex. It also establishes a geographical information system (GIS) of the surrounding 70 villages. Warana web-server gives market arrival and the daily prices of various regulated commodities. It includes the management information system for sugarcane cultivation developed by the NIC.

*Aiming at community development at large*

*i-Villages of Pondicherry of Swaminathan Research Foundation:* The MSSRF's (MS Swaminathan Research Foundation, Chennai) Information (i)-Villages are among the oldest examples of ICT initiatives in rural India aimed at ICT-enabled knowledge delivery to the poor for alleviating rural poverty. It is a hub and spikes model. The main hub with VSAT internet connectivity is linked to 20 village knowledge centers through a hybrid wired and wireless network consisting of PCs, telephones, VHF duplex radio and email through dial-up telephone lines. The i-Villages of Pudicherry are a good demonstration of the catalyzing role ICTs can play in capacity building, empowerment and raising incomes in rural areas,

*i-Community of Hewlett-Packard:* The i-Community of Hewlett-Packard (HP) at Kuppam in Andhra Pradesh provides an interesting contrast to the *e-Choupal c*ase in the business model adopted. HP is not an agribusiness company like ITC but a core ICT components and knowledge company. Its initiative represents an experiment (living lab) to become a partner in community development to gain knowledge and contacts that will make the company a stronger competitor in the emerging global knowledge economy. A Community

*Providing IT services in villages--business model*

*n-Logue:* India has over 740 million villagers living in over 6 38 000 villages. The villagers cannot afford to procure computer, internet and related products to benefit from recent IT-technology. The IIT, Chennai demonstrated through n Logue as to how to create rural services organization which would work towards delivering relevant and cost-effective technologies to the rural areas, which can be used to improve the living standards of Indian villages. IIT, Chennai set up the Telecommunications and Computer Networks (TeNeT)

Group. The group used CorDECT WLL (Wireless in Local Loop). CorDECT technology, jointly developed by TeNet and Midas Communication Technologies Pvt. Ltd, is much cheaper than the conventional telephone technologies and also provides greater bandwidth (35-70 kbps) than the dial up connection. The organization n-Logue was created to serve the information and communications needs of people living in small towns and rural areas of India. Through n-Logue, TeNeT intends to deliver ICT applications for essential services related to education, agriculture, health care and e governance through setting up a network of village internet centres called "Village Kiosks'. The kiosks provide 3 kinds of services: basic telephone service, Internet and basic computer service. The kiosk consists of a cor DECT (an advanced wireless access system) having wall-set with its accessories, a telephone and a telephone meter, a multimedia computer, power backup and Indian languages software. The local service provider (LSP) takes the whole set from IT for Rs 50 000. Several people are already making profits from their kiosks.

The company, 'n-Logue Communications Pvt Ltd', according to Prof. Ashok Jhunjhunwala of IIT, Chennai, is poised to bring about public call offices (PCOs) kind of revolution. However, in place of a PCO, kiosks are used. To rapidly scale its operations, the company employs a three tiered business model. These three tiers are:

(i) At the top level is n-Logue, which provides equipment, training and support to the Local Service Providers (LSPs) and kiosks, and also takes care of regulatory and connectivity issues. (ii) At the second level, n-Logue identifies and partners with a local entrepreneur (Local Service Provider or LSP) in each area it wishes to operate. LSPs find subscribers, provide services and collect payments. At the bottom level are the village kiosks, which provide services and information aimed at the rural market. With the help of n-Logue, the LSPs recruits the local entrepreneurs who set up the kiosks. n-Logue has successfully demonstrated as to how technology can be converted into sustainable business model and how building organizations towards this end can be achieved.

State Government of Madhya Pradesh based in a relatively remote and tribal dominated Dhar district of the State. "Gyandoot literally means messenger or purveyor of knowledge.

It was commissioned on 1 January 2000 beginning with the new millennium, as part of an exploration of e-governance. It is a low-cost, rural and community-owned intranet-based Government to Citizen (G2C) service delivery portal, managed by the district administration, which is connected to 39 village level community information centres through telephone dial up or wireless access. "Gyandoot' aims at improving governance at the village level by offering a range of services related to rural development that include information on weather, market prices and crop management (particularly soybean, wheat and gram) for the farmers. Kiosks are largely owned by local community and operated by an entrepreneur selected by the village committee. This entrepreneur pays 10% of his income to district administration for maintenance of the network. On the other hand, kiosks can also be owned privately and the private owner has to pay a fee of Rs 5000/year to the G*yandoot Samiti*. On an average about 2 340 farmers presently access the 'Gyandootportal every week and about 10 farmers visit a centre each day to access the services. A recent World Bank sponsored evaluation of the Project by the Centre for Electronic Governance, Indian Institute of Management, Ahmedabad has found that the project has succeeded in generating awareness of ICT among rural communities and reduced corruption and harassment.

*Other rural-based initiatives*

Other Indian initiatives involving ICTs for direct agricultural and rural concerns in India could be listed as:

* Knowledge Network for Grass Root Innovations:

   Several organizations such as the National Innovation Foundation (NIF) and the Honey Bee Network under the Society for Research and Initiatives for Sustainable Technologies and Institution (SRISTI) have been maintaining the database and national registers for green grassroots' technological innovations and traditional knowledge. These programmes insist on building of value chain of grassroot innovations including benefit sharing (Gupta 1999, 2001, 2003). Traditional Knowledge Digital Library (TKDL) is an Indian initiative, aims at documenting and classifying according to International Patent Classification (IPC), knowledge from traditional systems. The TKDL by the GOI is a welcome step to authenticate traditional knowledge at par with industrial property systems. This facility has evolved a scientific classification approach named as Traditional Knowledge Resource Classification (TKRC), which would enable retrieval of information on traditional knowledge in a scientific and rational manner. Its linkage to International Patent Classification would help patent examiners at the

*Facilitating e-Governance*

e-Governance consists of continuous optimization of service delivery, constituency participations and governance by transforming internal and external relationships through technology, internet and new media. It seeks to achieve efficiency, transparency and citizens' participation. It, thus contributes towards good governance, trust and accountability, citizen's awareness and empowerment, nation's growth, citizen's welfare and upholding democracy at large.

*'Gyandoot':* It is an e-governance initiative owned by the global level and increase patent examination substantially.

AGRICULTURE AND RELATED AREAS Application of Satellite Communication for Training Field Extension Workers in Rural Areas (Indian Space Research Organisation)

- The Andhra Pradesh State Wide Area Network (APSWAN) aims to link the state government with 23 district headquarters, serving as the backbone for 'multimedia services'. It is aimed to facilitate co-ordination between state headquarters and district offices in managing various regulatory, developmental and hazard mitigation programmes of the state government.

Automated Milk collection centres of Amul dairy cooperatives (Gujarat)

Centre for Alternative Agriculture Media (CAAM), a NGO initiative for promoting agriculture (Andhra Pradesh) and rural development.

- Computer-Aided Online Registration Department

- FRIENDS: It is single-window delivery mechanism of Kerala government services. The acronym FRIENDS stands for Fast, Reliable, Instant, Efficient, Network for Disbursement of Services. It is operational

OTHER INITIATIVES TAKEN IN E-LEARNING, E

*UN/FAO's initiatives*

The UN had several initiatives, particularly the Information and Communications Technologies Task Force and the Global Alliance for ICT and Development that have made a substantive contribution towards fostering an information society for agriculture and human welfare by sponsoring several regional meetings, organizing a series of global forums, producing several publications, linking developmental programmes of information and communication technology (ICT) with science and technology and the Millennium Development Goals and providing substantial input to the 'Partnership on Measuring ICT for Development (Ghosh 2007). The United Nations Food and Agriculture Organization (FAO) has provided an "e-Agriculture platform by launching a new interactive web-based site in 2007

UN's New Knowledge Management (KM) initiative also concerns knowledge management in food and nutrition community in India. The knowledge management initiative believes that India is one of the fastest growing economies and also has a vast knowledge base in several sectors. While some of this knowledge has been codified, shared and replicated, there is a large pool of knowledge that remains tacit. To harness this knowledge the United Nations country team in India launched a knowledge-sharing platform branded as "Solution Exchange which connects development professionals in similar fields from diverse organizations. It project is building Communities of Practice through electronic mail groups and also face-to-face interactions. Solution Exchange seeks to empower practitioners by offering them knowledge on demand-based on solutions from their peers. It is free, and impartial, demand-driven and solution-oriented service. It functions as a mailgroup with a moderating team. Problems and challenges are put as a query in an e-mail and posted to all community members. Members offer advice, experience, contacts or suggestions. A consolidated reply is prepared with a synopsis of original responses, additional resources and links. These are available on the website http://www.solutionexchange-un.net.in. E-discussion papers, newsletters, updates etc. are also made available as well as face-to-face meetings. So far the communities of practice are built around broad themes corresponding to the MDG goals and India's development priorities, viz ICT for development, poverty reduction, environment, health, gender, food and nutritional security, education and disaster management. Solution Exchange for the Food and Nutrition Security is a group of professionals from a wide range of organizations who are actively engaged in meeting the in Kerala with the involvement of poor women's groups. It promotes improved co-ordination between government and citizens in paying bills, obtaining applications, remitting registration fees, and so on.

Gramsat Pilot Project was launched by the Orissa Government to address governance issues, such as transparency, accountability, responsiveness, reduction of corruption, training and skill development, planning and monitoring, and people's participation. The project aims at creating a database that will include spatial data on land information, water geology, village and forest boundaries, river and drainage network and power distribution network.

Online Marketing and CAD in northern Karnataka; *Mahitiz-samuha* (Karnataka), a NGO initiative

• Raj Nidhi Kiosks: It operates in Rajasthan's Nyala village and is a web-enabled information kiosk system, designed and developed by the Department of Information Technology, Government of Rajasthan. These kiosks are meant to provide the citizens access to information related to agriculture, health, family planning, immunization schedules for children, employment, transportation, distance education, water and electricity connection, birth and death registration etc,

• VOICES – Madhyam Communications (Karnataka), a NGO initiative. Some exclusive agricultural initiatives for rural concerns include portals such as Haritgyan.com, Krishiworld.net,

country's food and nutrition security goals. The community promotes, among others (1) sustainable improvements in food security, (ii) reducing malnutrition and poverty, (iii) improved implementation and impact of food-related social safety net programmes, (iv) food safety and the prevention of food borne diseases, and (v) dietary diversification to prevent micronutrient deficiencies. Toeholdindia.com, Agriwatch.com, Acquachoupal.com, Plantersnet.com, etc. | 107 However, deepening the reach of the network, scaling up, barriers like literacy/e-literacy, information in vernacular languages and access to power, telephony and internet, bringing in cultural transformation in knowledge sharing, are some of the key challenges (Ghosh 2007).

*ICT initiatives in agricultural research and education*
The Indian Council of Agricultural Research (ICAR) is promoting E-learning and use of ICTs, particularly through the National Agricultural Innovation Project (NAIP).
(i) Efforts towards setting up a 'Secured intranet and
Central Data Centre for NARS: involving about 300 points under NARS have started. The project is to be
implemented by the ERNET. (ii) Developing the e-courses for BSc (Agri) and BVSC
degree programmes is envisaged. (iii) A digital library of PhD theses is being set up and it is proposed to digitize the PhD thesis submitted since
2000. (iv) The consortium for e-Resources in Agriculture (CERA)
is set up at IARI, New Delhi for providing access to e journals and e-resources to about 120 NARS libraries. A proposal on knowledge management in agriculture through the use of recent ICT tools and techniques in a consortium mode involving ICRISAT, IITs and ICAR institutes and SAUs is at the advanced stages of
approval. It is contemplated to provide e-connectivity to ICAR Units, KVKs and
Agricultural Universities (AUS). It comprises (1) e-Connectivity of KVKS (KVK-net), (2) ICAR Data center (DC) (ICAR-net), and (3) video Conferencing among ICAR
Institutes/AUs. In this regard, a component for linking of ERNET's Delhi point-of-presence (POP) to the National Agricultural Science Centre (NASC) control centres is integrally built-in. In case of Video conferencing, connectivity from ERNET POP at New Delhi and ICAR's videoconferencing gateway at NASC has been established. The three facilities are envisaged to be appropriately integrated
(V) AP
ultra-high speed CORE (multiples of 10 gigabit/second), complimented with a distribution at appropriate speeds. The participating institutions at the Edge shall connect to the National Knowledge Network seamlessly at speeds exceeding 1 gigabits/second or higher and the network architecture and governance structure shall allow the user institutions an option to connect to the distribution layer through a self-arranged/-procured last mile connectivity bandwidth. The main emphasis in the iNKN will be strong and robust internal-Intra-Net connectivity (and not ISP activities), so that India is seen and felt as one country from Himalayas to Kanyakumari.

Applications, such as countrywide classrooms will mitigate the systemic difficulties faced by planners for quality faculty and enhance the reach of all levels of education cutting across all barriers, geographic, religion, caste, economic, etc. The crux of the success of the Knowledge Network is related to the education-related applications, databases and delivery of these to the education users on demand.

The content creation and content sharing in all areas of science and technology including biotechnology and nanotechnology will be enabled for the purpose of education and research. The iNKN is proposed to be implemented through ERNET, an autonomous body.

INTELLECTUAL PROPERTY RIGHT ISSUES

Providing information and knowledge is fraught with competition and protection of rights. It has to keep pace with emerging provisions of related regulatory mechanisms and rapidly developing legal ramifications thereof, Intellectual property law raises issues of patent, copyright, and software infringement, as well as issues of institutional trademark. Fair use doctrine suggests that using materials exempt from copyright law in the classroom may be unlawful online.

In the present agricultural education scenario, it is, nevertheless, heartening to note that several information systems are being built. Fortunately, there are suppliers who offer electronic versions of information including copyrighted material over the Internet. This material generally consists of electronic journals with multimedia supplements, electronic reference works, electronic books, and books with electronic components etc. There is Intellectual Property Right issues involved and settled in regard to access and permitted use. These suppliers enter into agreement with the licensee who in turn has authorized users mainly in the form of faculty members of the universities, students and staff members. The license is generally a non-exclusive one granted to the licensee to use the licensed material and to provide the licensed material to authorized users via licensee's secure network. Licensed materials are generally protected by (i) copyright, and/or (ii) database rights. All rights not specifically granted to licensee are expressly reserved. There is, however, a way of providing public access, i e through workstations on library premises

*Integrated National Knowledge Network to connect all*

*knowledge institutions*

Establishment of Integrated National Knowledge Network (iNKN), (to be called "Viswaroopam), is being vigorously pursued as a national priority and national policy. The objective of the INKN is to bring together all the stakeholders in health, agriculture, science and technology, higher education, grid computing and research and development. The architecture of the Integrated National Knowledge Network will be scalable and the network will consist of an and not through remote access. If the licensee provides public access to its library collection, it may also provide access to and permit copying from the licensed materials by members of the public for their scholarly research and personal use from workstations on library premises. Any form of remote access to the licensed materials by members of the public is not permitted.

Despite these, fair use of material and content may be allowed to go unrestricted. The medium and mode itself should not make otherwise unrestricted material as a restricted one.

FUTURE STRATEGY--SUGGESTIONS Among the biggest challenges to the integration and diffusion of ICT facing developing countries are insufficient policy, strategies and implementation capacity. Many current policies, strategies and practices, designed for site-based education and print-on-paper resources, need a re-look, as those may be inappropriate or insufficient in an online environment. ICT should move closer to the mainstream of development economics and policies, nationally, regionally and globally. However, movement in this respect has been slow. Some of the strategic measures that converge towards 'information society' strategies are:

• Institutional networking and formation of consortia

appears to be necessary to promote knowledge-sharing using e-Learning. This would include common e-Learning platforms in agricultural universities. Consortia may also help in making feasible suggestions to reshape policies and practices, particularly for fostering a distributed education environment. For example, state regulation and regional accreditation policies may require change. Leaving aside the macro level policies, the consortia may have such inter- and intra-institutional changes effected that are commonly agreed and mutually beneficial. For example, institutions may need to modify faculty policies on workload, class size, and compensation. Students who are low-income, less privileged, less well-prepared, or who have disabilities may need guaranteed access to appropriate technologies, services, and financial assistance.

- The Government may need to amend certain existing

policies that do not yet provide or prohibit governmental aid to students in many distance-learning programme. Content development, connectivity, capacity building and other desiderata are needed to be strengthened and more institutionalized to integrate ICTs into the process of agricultural development in the country. The grand Indian initiatives such as iNKN are to be highly appreciated. Nevertheless, local area networks have to supplement this highway. Rao (2007) concluded that for a reliable and effective last mile connectivity, VSAT is technically the best, though expensive. Local access solutions like n-Logue provide alternative means to relatively inexpensive broadband connectivity directly from the village centre level. The next effective option is the hub and spokes model of MSSRF Self help group or NGO-led ICT-based models may need (supplemental) funding from external agencies to remain viable in the long run. The most desirable thing to happen would be (i) a suitable blend of community services with profit motive in private sector led endeavours, and (ii) needed flexibility and economic viability in the public sector initiatives. It is often felt that ICT initiatives in rural India cannot be extensively taken owing to high cost. Several models have flayed this notion. For example, *Gyandoot*, with an investment of only Rs 2.5 million, could make an impact on the stakeholder community, showing that such initiatives need not be investment intensive (Meera *et al.* 2004).

- Partnership with leading IT institutions needs to be

forged and fostered towards developing the content as per global standards based on the instructional technology and technical content provided by the scientists and teachers. The ICT initiatives need to strengthen e-Learning and digital multimedia resource development and simultaneously promote the use of this media to offer distance learning. Policy for promotion of open sources in content creation, deployment and management should further facilitate the cause both in "presence and distance mode. Policy or incentives for private sector as well as NGOs are needed towards establishing and maintaining information kiosks at district and block level. For community services, the systems may provide free access to learning to the users in the region. The bandwidth and hardware and infrastructure for such welfare measures may be provided by the government not only to public sector and NGOs but also to encourage private sector participation so that the rural and per-urban youth are enabled to climb the skill ladder faster and eventually move towards entrepreneurship and gainful self-employment. In conclusion, effective utilization of ICT for knowledge sharing and resorting to e-Learning in education has the potential to make the agriculture sector and rural communities in India prosperous. This potential has started being realized in several cases. There are models, spread across different segments of agriculture for convergence of ICT with agricultural development that can be emulated, hybridized, and improvised to bring about farm-prosperity at large. Several desiderata, nevertheless, are to be further implemented. It is not a challenge that Governments can overcome alone; the private sector and civil society bring unique assets to the fore. For ICT initiatives to be successful and sustainable in the long run, collaborative efforts are indispensable.

ACKNOWLEDGEMENTS

Author is grateful to all the respondents who cared to send their views and literature as hard copies as well as to those who discussed telephonically, indebted to the able team of social scientists of the National Centre for Agricultural Economics and Policy Research (NCAP), ICAR, New Delhi who provided valuable suggestions to improve the manuscript, and grateful to Google India Pvt. Ltd for all the support and encouragement provided.

REFERENCES

Adhiguru P and Mruthyunjaya. 2004. Institutional innovations for
    using Information and Communication Technology, Policy Brief 18, National Centre for
                    Agricultural Economics and Policy
Research, New Delhi. CEC. 2001. Communication from the Commission to the Council and the European Parliament. The e-Learning Action Plan: Designing tomorrow's educatiion. (http://*europa.eu.int/comm*/
education/elearnin*g*/index.html). Ghosh and Gopi N. 2007. UN's New KM Initiative: Knowledge
  Management in Food and Nutrition Community in India. (Communication from Assistant
        FAO Representative & Resource Person, Food & Nutrition Security Community SOLUTIONEXCHANGE). Gilder G. 1993. Metcalfe's Law and legacy. *Forbes* ASA*P*, Sept. 13, 1993. GoI. 2001. Report of Prime Minister's Task Force on India as
Knowledge Superpower. Planning Commission, Government
of India, New Delhi. Gupta A K. 1999. Conserving Biodiversity and Rewarding Associated Knowledge and Innovation Systems: Honey Bee Perspective. Paper presented at First Commonwealth Science Forum-Access, Bioprospecting, Intellectual Property

Rights and Benefit Sharing and the Commonwealth, Goa, 23–25 September 1999. Gupta A K. 2001. Framework for rewarding indigenous knowledge in developing countries: Value chain for grassroots innovations, *Paper presented at* W*TO Expert Committee,* beld on 3 September 2001. Gupta AK. 2003. Rewarding conservation of biological and genetic resources and associated traditional knowledge and contemporary grassroots creativity. IIM, Ahmedabad W.P. No. 2003-01-06, January 2003.

"*Gyandoot',* Gyandoot Bulletin, District Administration, Dhar, Madhya Pradesh. I-kisan Limited. 2000. *ikisan,* folder, Ikisan Limited. ITC's eChoupal Initiative: http://www.itcibd.com/e-choupall.asp Jayaraman K S. 2002. India Online. N*ature* 415: 358-9. Lobo, Albert and Balakrishnan, Suresh (2002) Report Card on Service of Bhoomi Kiosks: An assessment of benefits by users of the computerized land records system in Karnataka. Public Affairs Center Bangalore. Bangalore, India and the World Bank, November. pp 1-10. Retrieved from: http://unpan1.un.org/intradoc/groups/public/documents/APCITY/UNPANO 15135.pdf Maru, Ajit. 2003. ICT Status - Asia Pacific, in *ICT Regional Status Report 2003.* APAARI Publication. Meera Shaik N.2002. 'A Critical analysis of information technology in agricultural development: Impact and implications."

Unpublished PhD thesis, IARI, New Delhi 110 012. Meera Shaik N, Jhamtani A and Rao DUM. 2004. Information and communication technology in agricultural development: A comparative analysis of three projects from India. ODI, Agricultural Research and Extension Network (AGREN) *Network Paper No.135.* pp 1-14. Metcalfe B. 1995. Metcalfe's Law: A network becomes more valuable as it reaches more users. Infoworld, Oct. 2, 1995. See also the May 6, 1996 column, (http://www.infoworld.com/cgi bin/displayNew.pl?/metcalfe/bm 050696.htm). NIC. 2005. *Good Governance through ICT.* pp 1-153. Shefali S Dash *et al.* (Ed). National Informatics Centre (GoI) Publication. Nambiar M. Madhavan. 2005. ICT for Education: The Experience of India. *(in) Harnessing the Potential of ICT for Education -* A *Multistakeholder Approach.* pp 19-21. Bonnie Bracey and Terry Culver (Eds). Proceedings from the Dublin Global Forum of the United Nations ICT Task Force. Ed. UN Publication. Orbicom 2004. India - Overview. Madanmohan Rao. pp 107–13. http://www.digital-review.org/2005-6PDFs/2005%20C09%20in%20India% 20107-113.pdf. Rao N H. 2007. A framework for implementing information and communication technologies in agricultural development in India. *Technology Forecasting and Social Change* 74:491-518. Seth D P S. 2006. National Knowledge Commisssion -- A Report on National Knowledge Network (Dec. 2006). http:// www.knowledgecommission, gov.in/downloads/documents/er_nknet.pdf. Succi C and Cantoni L. 2005. Quality Benchmarking for eLearning in European Universities. *(in) Proceedings of World Conference on Educational Multimedia, Hypermedia and Telecommunications 2005*. pp 116-23. P. Kommers and Richards G (Eds.). Chesapeake, VA: AACE. Swaminathan MS. 1993. (Ed.) *Information Technology: Reaching the Unreached*. Chennai, Macmillan India.